\documentclass{amsart}
\usepackage{amssymb}
\usepackage{a4}
\begin{document}
\newtheorem{theorem}{Theorem}[section]
\newtheorem{lemma}[theorem]{Lemma}
\newtheorem{remark}[theorem]{Remark}
\newtheorem{definition}[theorem]{Definition}
\newtheorem{corollary}[theorem]{Corollary}
\newtheorem{example}[theorem]{Example}
\def\BB{\mathcal{B}}
\def\ffrac#1#2{{\textstyle\frac{#1}{#2}}}
\def\id{\operatorname{Id}} \def\qedbox{\hbox{$\rlap{$\sqcap$}\sqcup$}}
\makeatletter
  \renewcommand{\theequation}{%
   \thesection.\alph{equation}}
  \@addtoreset{equation}{section}
 \makeatother
\title[Heat Content Asymptotics]
{Heat Content asymptotics for operators of Laplace type with
spectral boundary conditions} \author{P. Gilkey, K. Kirsten, and
JH. Park} \begin{address}{Mathematics Department, University of
Oregon, Eugene Or 97403 USA}
\end{address}
\begin{email}{gilkey@darkwing.uoregon.edu}\end{email}
\begin{address}{Department of Mathematics, Baylor University, Waco, TX 76798, USA}\end{address} \begin{email}{Klaus\_Kirsten@Baylor.edu}\end{email}
\begin{address}{Department of Computer and Applied Mathematics, Honam  University,
  Gwangju 506-714 Korea}\end{address} \begin{email}{jhpark@honam.ac.kr}\end{email}
\begin{abstract} Let $P$ be an operator of Dirac type and let $D=P^2$ be the associated operator of Laplace type.
We impose spectral boundary conditions and study the leading heat
content coefficients for $D$.
\end{abstract} \keywords{Laplace type, Dirac type, heat content
asymptotics, spectral boundary conditions}
\subjclass[2000]{Primary 58J50} \maketitle

\section{introduction}\label{sect1} Let $P$ be an operator of Dirac type on a vector
bundle $V$ over a compact Riemannian manifold $M$ of dimension $m$ with smooth boundary
$\partial M$. Let $D:=P^2$ be the associated operator of Laplace type. The leading symbol $\gamma$ of $P$
defines a Clifford module structure on $V$. Choose an auxiliary connection $\nabla$ on $V$ so that
$\nabla\gamma=0$. Adopt the Einstein convention and sum over repeated indices; indices $i,j$
will range from $1$ to $m$ and index a local orthonormal frame $\{e_i\}$ for $TM$.
Expand $$P=\gamma_i\nabla_{e_i}+\psi_P\,.$$

We must impose suitable boundary conditions. As $P$ need not admit
local boundary conditions, we shall consider spectral boundary
conditions; these were first introduced by Atiyah et. al.
\cite{atiy75-77-43} to study the index theorem for manifolds with
boundary. Near the boundary, normalize the local frame so that
$e_m$ is the inward unit geodesic vector field. Let indices $a,b$
range from $1$ to $m-1$ and index the induced orthonormal frame
for $T\partial M$. Let $$A=-\gamma_m\gamma_a\nabla_{e_a}+\psi_A$$
for some endomorphism $\psi_A$ of $V|_{\partial M}$; $A$ is of
Dirac type on $V|_{\partial M}$ with respect to the induced
tangential Clifford module structure
$\gamma^T_a:=-\gamma_m\gamma_a$.

For the sake of simplicity, we shall assume $A$ has no purely
imaginary eigenvalues. Let $\Pi_A^+$ be spectral projection on the
span of the generalized eigenspaces of $A$ corresponding to
eigenvalues with {\bf positive} real part. This spectral
projection defines a boundary condition for $P$; the associated
boundary operator for $D$ is $$\BB:=\Pi_A^+\oplus\Pi_A^+P\,.$$ Let
$D_\BB$ be the associated realization. The fundamental solution
$e^{-tD_\BB}$ of the heat equation is well defined;
$u=e^{-tD_\BB}\phi$ is characterized by the relations
$$(\partial_t+D)u=0,\quad\BB u=0,\quad\text{and}\quad
u|_{t=0}=\phi\,.$$ We refer to Grubb
\cite{grub92-17-2031,grub99-37-45} and Grubb and Seeley
\cite{grub93-317-1123,grub95-121-481,grub96-6-31} for further
details.

Let $\langle\cdot,\cdot\rangle$ be the natural pairing between $V$
and the dual bundle $V^*$, let $dx$ be the Riemannian measure on
$M$, and let $\rho\in C^\infty(V^*)$ be the specific heat. The
total heat energy content of the manifold with initial temperature
$\phi$ is given by
$$\beta(\phi,\rho,D,\BB)(t):=\textstyle\int_M\langle
u(x;t),\rho(x)\rangle dx\,.$$

If we had imposed local boundary conditions such as Dirichlet or
Robin, then it is well known that there is a complete asymptotic
series for $\beta$ with locally computable coefficients $\beta_n$.
We refer to
\cite{berg93-2-147,berg94-120-48,desj94-215-251,gilk02-59-269,
avit93-26-823,macme03-200-150,sav98-73-181,sav01-288-432} for a
discussion of this case. Thus we assume there exists a complete
asymptotic series as $t\downarrow0$ of the form
$$\beta(\phi,\rho,D,\BB)(t)\sim\sum_{n=0}^\infty\beta_n(\phi,\rho,D,\BB)t^{n/2}\,.$$

If $\Psi$ is an operator on $C^\infty(V)$ or on $C^\infty(V^*)$,
let $\tilde\Psi$ be the formal adjoint on $C^\infty(V^*)$ or on
$C^\infty(V)$, respectively. Similarly, if $\Psi$ is an operator
on $C^\infty(V|_{\partial M})$ or on $C^\infty(V^*|_{\partial
M})$, let $\tilde\Psi$ be the adjoint on $C^\infty(V^*|_{\partial
M})$ or on $C^\infty(V|_{\partial M})$, respectively. Let
$\tilde\nabla$ be the dual connection on $V^*$, let $L_{ab}$ be
the second fundamental form of the boundary, and let $dy$ be the
Riemannian measure on the boundary. We omit the proof of the
following Lemma in the interests of brevity as it is
straightforward.

\begin{lemma}\label{lem-1.1}
\begin{enumerate}
\item $\tilde P=-\tilde\gamma_i\tilde\nabla_{e_i}+\tilde\psi_P$ and $\tilde A=-\tilde\gamma_m\tilde\gamma_a\tilde\nabla_{e_a}+\tilde\psi_A$.
\item Let $A^{\#}:=\tilde\gamma_m\tilde A\tilde\gamma_m$ on $C^\infty(V^*|_{\partial M})$. Then: \begin{enumerate}\item  $\Pi_{A^{\#}}^+$ defines the adjoint boundary condition for $\tilde P$. \item $\tilde B:=\Pi_{A^{\#}}^+\oplus\Pi_{A^{\#}}^+\tilde P$ defines the adjoint boundary condition for $\tilde D:=\tilde P^2$. \item Let $\psi_{A^\#}:=\tilde\gamma_m\tilde\psi_A\tilde\gamma_m+L_{aa}\id$. Then
   $A^{\#}=-\tilde\gamma_m\tilde\gamma_a\tilde\nabla_{e_a}+\psi_{A^{\#}}$.
\end{enumerate}
\item
$\textstyle\int_M\{\langle D\phi,\rho\rangle-\langle\phi,\tilde
D\rho\rangle\}dx= -\textstyle\int_{\partial
M}\{\langle\gamma_m\Pi_A^+P\phi,\rho\rangle +\langle
P\phi,\tilde\gamma_m\Pi_{A^{\#}}^+\rho\rangle$\par$
+\langle\phi,\tilde\gamma_m\Pi_{A^{\#}}^+\tilde P\rho\rangle
+\langle\gamma_m\Pi_A^+\phi,\tilde P\rho\rangle\}
 dy$.
\item We have $(\id-\widetilde{\Pi_A^+})\tilde\gamma_m=\tilde\gamma_m\Pi_{A^{\#}}^+$.
\item If $\BB\phi=0$ and if $\tilde\BB\rho=0$, then $\int_M\{\langle D\phi,\rho\rangle-\langle\phi,\tilde D\rho\rangle\}dx=0$. \end{enumerate} \end{lemma}

We can now state the main result of this paper:
\begin{theorem}\label{thm-1.2} Adopt the notation established
above. \begin{enumerate} \item
$\beta_0(\phi,\rho,D,\BB)=\textstyle\int_M\langle\phi,\rho\rangle
dx$. \item
$\beta_1(\phi,\rho,D,\BB)=-\frac2{\sqrt\pi}\textstyle\int_{\partial
M}\langle\Pi_A^+\phi,\Pi_{A^{\#}}^+\rho\rangle dy$.
\item $\beta_2(\phi,\rho,D,\BB)=-\textstyle\int_M\langle D\phi,\rho\rangle dx
+\textstyle\int_{\partial
M}\{-\langle\gamma_m\Pi_A^+P\phi,\rho\rangle-\langle\gamma_m\Pi_A^+\phi,\tilde
P\rho\rangle $\par\qquad$
+\frac12\langle(L_{aa}+A+\widetilde{A^{\#}}-\gamma_m\psi_P
+\psi_P\gamma_m-\psi_A-\widetilde{\psi_{A^{\#}}})\Pi_A^+\phi,\Pi_{A^{\#}
}^+\rho\rangle\}dy$.
\end{enumerate}
\end{theorem}

A variant of this result was established in \cite{gilk-5-49} using
a special case calculation. In this paper, we extend this result
by using functorial methods. In Section \ref{Sect2}, we derive
various naturality properties of these invariants. These
properties are then used in Section \ref{Sect3} to complete the
proof.

\section{Properties of the heat content invariants}\label{Sect2}\

We begin with some general observations:

\begin{lemma}\label{lem-2.1}
\begin{enumerate}
\item $\beta_0(\phi,\rho,D,\BB)=\textstyle\int_M\langle\phi,\rho\rangle dx$.
\item $\beta_n(\phi,\rho,D,\BB)=\beta_n(\rho,\phi,\tilde D,\tilde\BB)$. \item If
$\BB \phi=0$, then $\frac n2\beta_n(\phi,\rho,D,\BB)= - \beta_{n-2}(D\phi,\rho,D,\BB)$.
\item If $M$ is closed, then $\beta_n(\phi,\rho,D,\BB)=0$ if $n$ is odd while if $n=2k$ is even,
then $\beta_{2k}(\phi,\rho,D,\BB)=(-1)^k\frac1{k!}\textstyle\int_M\langle D^k\phi,\rho\rangle dx$.
\item $\beta_n(\phi,\rho,(-P)^2,\BB)=\beta_n(\phi,\rho,P^2,\BB)$.
\end{enumerate}\end{lemma}

\begin{proof} The first assertion is immediate since $u|_{t=0}=\phi$.
To prove the second assertion, we set $u(x;t):=e^{-tD_\BB}\phi$ and $\tilde u(x;t):=e^{-t\tilde D_{\tilde\BB}}\rho$.
We then have $$\partial_tu=-Du\quad\text{and}\quad\partial_t\tilde u=-\tilde D\tilde u\,.$$
Fix $t>0$. For $0\le s\le t$, set $f(s):=\textstyle\int_M\langle u(x;s),\tilde u(x;t-s)\rangle dx$. By Lemma \ref{lem-1.1} (5), \begin{eqnarray*} \partial_sf(s)&=&
  \textstyle\int_M\{\langle \partial_su(x;s),\tilde u(x;t-s)\rangle
    -\langle u(x;s),\partial_t\tilde u(x;t-s)\rangle\} dx\\
    &=&\textstyle\int_M\{-\langle Du(x;s),\tilde u(x;t-s)\rangle
   +\langle u(x;s),\tilde D\tilde u(x;t-s)\rangle\}dx\\
&=&0
\end{eqnarray*}
since $\BB u=0$ and $\tilde\BB\tilde u=0$. Because $f(s)$ is
constant, Assertion (2) follows as \begin{eqnarray*}
0&=&f(t)-f(0)=\textstyle\int_M\{\langle u(x;t),\tilde
u(x;0)\rangle
   -\langle u(x;0),\tilde u(x;t)\rangle\}dx\\ &=&\beta(\phi,\rho,D,\BB)(t)-\beta(\rho,\phi,\tilde D,\tilde\BB)(t). \end{eqnarray*}

To establish the third assertion, we suppose that $\BB\phi=0$.
Since $\tilde\BB\tilde u=0$, we can use Lemma \ref{lem-1.1} (5)
and Assertion (2) to see \begin{eqnarray*}
&&\partial_t\beta(\phi,\rho,D,\BB)(t)
=\partial_t\beta(\rho,\phi,\tilde D,\tilde\BB)(t)\\
&=&\textstyle\int_M\langle\partial_t\tilde u(x;t),\phi(x)\rangle
dx =-\textstyle\int_M\langle \tilde D\tilde u(x;t),\phi(x)\rangle
dx\\ &=&-\textstyle\int_M\langle\tilde u(x;t),D\phi(x)\rangle dx
=-\beta(\rho,D\phi,\tilde D,\tilde\BB)(t)\\
&=&-\beta(D\phi,\rho,D,\BB)(t)\,. \end{eqnarray*} We equate terms
in the asymptotic expansions to derive Assertion (3). Assertion
(4) follows by induction from Assertions (1) and (3); Assertion
(5) is immediate. \end{proof}

Mixed boundary conditions will play an important role in our
discussion. Assume given an endomorphism $\gamma_0$ of
$V|_{\partial M}$ so that $\gamma_0^2=\id$. Let
$$\Xi_\pm:=\textstyle\frac12(\id\pm\gamma_0)\quad\text{and}\quad
    V_\pm:=\Xi_{\pm}\{V|_{\partial M}\}$$
be the associated spectral projections and eigenspaces. For
$S\in\operatorname{End}(V|_{\partial M})$, set
$$
\BB_{\gamma_0,S}\phi:=\Xi_-(\nabla_{e_m}+S)\phi|_{\partial
M}\oplus\Xi_+\phi|_{\partial M}\,. $$ The operator $D$ determines
a natural connection $\nabla^D$. The following Theorem, after
taking into account our sign conventions, follows from results of
\cite{desj94-215-251}.

\begin{theorem}\label{thm-2.2}
Adopt the notation established above. Then: \begin{enumerate}\item
$\beta_{0}(\phi,\rho,D,\BB_{\gamma_0,S})=
\int_{M}\langle\phi,\rho\rangle dx$. \item
$\beta_{1}(\phi,\rho,D,\BB_{\gamma_0,S})=-\ffrac2{\sqrt\pi}
\int_{\partial M}\langle\Xi_+\phi,\tilde\Xi_+\rho\rangle dy$.
\item
$\beta_{2}(\phi,\rho,D,\BB_{\gamma_0,S})=-\int_{M}\langle
D\phi,\rho\rangle dx+ \int_{\partial
M}\{\langle\Xi_-(\nabla_{e_m}^D\phi+S\phi),\rho\rangle
$\par\qquad\qquad\qquad$
+\langle\ffrac12L_{aa}\Xi_+\phi,\rho\rangle
-\langle\Xi_+\phi,\tilde\nabla_{e_m}^D\rho\rangle\} dy$.
\end{enumerate}\end{theorem}

\begin{example}\label{exm-2.3}
\rm We can relate spectral and mixed boundary conditions in the
following special setting. Let $(\theta_1,...,\theta_{m-1})$ be
the usual periodic parameters on the torus $\mathbb{T}^{m-1}$ and
let $r$ be the radial parameter on the interval $[0,1]$. Let $f\in
C^\infty[0,1]$ with $f(0)=f(1)=0$. Take a warped product on
$M:=\mathbb{T}^{m-1}\times[0,1]$ of the form:
$$ds_M^2:=e^{2f(r)}d\theta^a\circ d\theta^a+dr^2\,.$$ The volume
element is then given by $dx=gdrd\theta_1...d\theta_{m-1}$ where
$g:=e^{(m-1)f}$. Let $\varepsilon(0):=+1$ and $\varepsilon(1):=-1$
so that the inward unit normal is $\varepsilon\partial_r$. Let
$\Theta_i\in M_\ell(\mathbb{C})$ satisfy the Clifford commutation
relations $\Theta_i\Theta_j+\Theta_j\Theta_i=-2\delta_{ij}$. Set
$$\gamma_m:=\Theta_m,\quad\gamma_a:=e^f\Theta_a,
\quad\text{and}\quad\gamma^a:=e^{-f}\Theta_a\,.$$ Let $\gamma_0\in
M_\ell(\mathbb{C})$ satisfy $\gamma_0^2=\id$ and
$\gamma_0\gamma_m+\gamma_m\gamma_0=0$. Let $\delta_1>0$ and
$\delta_2>0$ be real parameters. Let $V=M\times\mathbb{C}^\ell$,
let $\phi=\phi(r)$ and let $\rho=\rho(r)$. Set \begin{eqnarray*}
&&P:=\gamma_m\partial_r+\gamma_a\partial_a^\theta
   +\delta_1\gamma_m\gamma_0\quad\text{on}\quad C^\infty(V),\\ &&A:=-\varepsilon\gamma_m\gamma_a\partial_a^\theta+\delta_2\gamma_0
\qquad\phantom{..A}\text{on}\quad C^\infty(V|_{\partial M})\,.
\end{eqnarray*} For generic values of $\delta_2$, $\ker(A)=\{0\}$.

Let $V_0:=[0,1]\times\mathbb{C}^\ell$. Let $\Xi_\pm$ be projection
on the $\pm1$ eigenspaces of $\gamma_0$; we have that
$\gamma_m\Xi_-=\Xi_+\gamma_m$. Let
$P_0:=\gamma_m\partial_r+\delta_1\gamma_m\gamma_0$ on
$C^\infty(V_0)$, and let $D_0:=P_0^2$. Set
$$\BB_0\phi:=\{\Xi_+\phi\oplus\Xi_+P_0\phi\}|_{\partial[0,1]}\quad\text{and}\quad
S:=\delta_1\gamma_0\,.$$
We show that $\BB_0$ and $\BB_{\gamma_0,S}$ define the same boundary conditions for $D_0$ by checking: $$\begin{array}{llll} &\Xi_+(\phi|_{\partial[0,1]})=0&\text{and}&\Xi_+P_0\phi|_{\partial[0,1]}=0,\\
\Leftrightarrow&\Xi_+(\phi|_{\partial[0,1]})=0&\text{and}&
\Xi_+\gamma_m(\partial_r\phi+\delta_1\gamma_0\phi)|_{\partial[0,1]}=0,\vphantom{\vrule
height 11pt}\\
\Leftrightarrow&\Xi_+(\phi|_{\partial[0,1]})=0&\text{and}&
\gamma_m\Xi_-(\partial_r\phi+\delta_1\gamma_0\phi)|_{\partial[0,1]}=0,\vphantom{\vrule
height 11pt}\\
\Leftrightarrow&\Xi_+(\phi|_{\partial[0,1]})=0&\text{and}&
\Xi_-(\partial_r+S)\phi|_{\partial[0,1]}=0\,.\vphantom{\vrule
height 11pt} \end{array}$$ \end{example}

\begin{lemma}\label{lem-2.4} $\beta_n(\phi,g^{-1}\rho,D,\BB)=(2\pi)^{m-1}\beta_n(\phi,\rho,D_0,\BB_{\gamma_0,S})$ in Example \ref{exm-2.3}.
\end{lemma}

\begin{proof}
Let $u_0:=e^{-tD_{0,\BB_0}}\phi$. Set $u(r,\theta;t):=u_0(r;t)$. If $\Phi=\Phi(r)$ and $\Psi=\Psi(r)$,
then $\Pi_A^+\Phi=\Xi_+\Phi$ and $\Pi_{A^{\#}}^+\Psi=\tilde\Xi_+\Psi$.
Thus we may show $u=e^{-tD_\BB}\phi$ by checking \begin{eqnarray*} &&(\partial_t+D)u=(\partial_t+D_0)u=0,\\
&&\BB u=\Pi_A^+u\oplus\Pi_A^+Pu=\Xi_+u\oplus\Xi_+P_0u=\BB_0u_0=0,\\
&&u|_{t=0}=u_0|_{t=0}=\phi\,.
\end{eqnarray*}
After taking into account the change in volume elements and
equating terms in the asymptotic expansions, the result follows.
\end{proof}

\begin{remark}\label{rmk-2.5}
\rm The coefficients appearing in Theorem \ref{thm-1.2} are
independent of the dimension $m$. This observation is quite
general. Let $P_0$ be an operator of Dirac type on a bundle $V_0$
over an $m$ dimensional manifold $M_0$. By doubling the rank of
$V_0$ and by replacing $P_0$ by $P_0\oplus-P_0$ if necessary, we
may suppose there exists $\gamma_0$ so
$$\gamma_0P_0+P_0\gamma_0=0\quad\text{and}\quad\gamma_0^2=-\id\,.$$
Define analogous structures on $M:=M_0\times S^1$ by setting
$$P:=P_0+\gamma_0\partial_\theta\qquad\text{and}\qquad
A:=A_0-\gamma_m\gamma_0\partial_\theta\,.$$ Let
$\phi(x,\theta)=\phi_0(x)$ and let $u_0:=e^{-tD_{0,\BB_0}}\phi_0$.
Set $u(x,\theta;t):=u_0(x;t)$. Then one has $u=e^{-tD_\BB}\phi$;
thus the formulae in dimension $m+1$ restrict to the corresponding
formulae in dimension $m$. By contrast, the corresponding formulae
for the heat trace asymptotics \cite{dowk99-242-107,gilk00u}
exhibit a very complicated dependence on $m$.
\end{remark}

\section{Proof of Theorem \ref{thm-1.2}}\label{Sect3}
Since $u|_{t=0}=\phi$, Assertion (1) of Theorem \ref{thm-1.2} is
immediate. The interior integrands in $\beta_1$ and $\beta_2$ are
determined by Lemma \ref{lem-2.1} (4). Thus we need only determine
the boundary integrands. We apply Lemma \ref{lem-2.1} throughout.
Dimensional analysis shows that the boundary integrands are
homogeneous of total weight $n-1$ in the jets of $\phi$, of
$\rho$, and of the derivatives of the symbols of $A$ and of $P$.
The spectral projections $\Pi_A^+$ and $\Pi_{A^{\#}}^+$ and the
endomorphisms $\gamma_i$ have weight $0$; the second fundamental
form $L$, the operators $A$ and $P$, and the endomorphisms
$\psi_A$ and $\psi_P$ have weight 1.

We begin by studying $\beta_1$; there is no interior contribution.
If $\BB\phi=0$, then by Lemma \ref{lem-2.1} (3),
$\beta_1(\phi,\rho,D,\BB)=0$. Dually by Lemma \ref{lem-2.1} (2),
$$\beta_1(\phi,\rho,D,\BB)=\beta_1(\rho,\phi,\tilde
D,\tilde\BB)=0\text{ if }\tilde\BB\rho=0\,.$$ Since the boundary
integrand for $\beta_1$ must be homogeneous of weight $0$, there
exist universal constants so \begin{equation}\label{eqn-3.a}
\beta_1(\phi,\rho,D,\BB)=\textstyle\int_{\partial
M}\{c_0(m)\langle\Pi_A^+\phi,\Pi_{A^{\#}}^+\rho\rangle
+c_1(m)\langle\gamma_m\Pi_A^+\phi,\Pi_{A^{\#}}^+\rho\rangle\}dy\,.
\end{equation}
By Lemma \ref{lem-2.1} (5),
$\beta_1(\phi,\rho,P^2,\BB)=\beta_1(\phi,\rho,(-P)^2,\BB)$.
Replacing $P$ by $-P$ replaces $\gamma_m$ by $-\gamma_m$. Thus
$\langle\gamma_m\Pi_A^+\phi,\Pi_{A^{\#}}^+\rho\rangle$ plays no
role so we may take $c_1(m)=0$.

We apply Lemma \ref{lem-2.4} with $f=0$. By Theorem \ref{thm-2.2}
and Equation (\ref{eqn-3.a}),
\begin{eqnarray*}
\beta_1(\phi,\rho,D,\BB)&=&(2\pi)^{m-1}\beta_1(\phi,\rho,D_0,\BB_0)
   =-\ffrac2{\sqrt\pi}\textstyle\int_{\partial M}\langle\Xi_+\phi,\tilde\Xi_+\rho\rangle dy\\ &=&c_0(m)\textstyle\int_{\partial M}\langle\Xi_+\phi,\tilde\Xi_+\rho\rangle dy\,. \end{eqnarray*} This shows that $c_0(m)=-\textstyle\frac2{\sqrt\pi}$ which completes the proof of Theorem \ref{thm-1.2} (2).

\begin{remark}\label{rmk-3.1}
\rm The constant $c_0(m)$ was determined in \cite{gilk-5-49} using
a special case computation and the present calculation should be
regarded as providing a useful cross check on that
calculation.\end{remark}

To prove the final assertion of Theorem \ref{thm-1.2}, we express
$\beta_2$ in terms of invariants with undetermined universal
coefficients.

\begin{lemma}\label{lem-3.2}
There exist universal constants $c_i$ so that
\begin{eqnarray*} &&\beta_2(\phi,\rho,D,\BB)=-\textstyle\int_M\langle D\phi,\rho\rangle dx+
  \textstyle\int_{\partial M}\{-\langle\gamma_m\Pi_A^+P\phi,\rho\rangle\\
&&\qquad-\langle\gamma_m\Pi_A^+\phi,\tilde P\rho\rangle
+\langle(c_2(A+\widetilde{A^{\#}})+c_3L_{aa}+c_4(\gamma_m\psi_P-\psi_P\gamma_m)\\
&&\qquad+c_5(\psi_A+\widetilde{\psi_{A^{\#}}}))
\Pi_A^+\phi,\Pi_{A^{\#}}^+\rho\rangle \}dy\,.
\end{eqnarray*}
\end{lemma}

\begin{proof} We argue heuristically. By Lemma \ref{lem-2.1},
the interior integral for $\beta_2$ is given by $-\langle
D\phi,\rho\rangle$. We define the normalized invariant
$\mathcal{C}$, which is given by a suitable boundary integral, by
the identity \begin{eqnarray*} \beta_2(\phi,\rho,D,\BB)&=&
\mathcal{C}(\phi,\rho,D,\BB)-\textstyle\int_M\langle
D\phi,\rho\rangle dx \\ && +\textstyle\int_{\partial
M}\{-\langle\gamma_m\Pi_A^+P\phi,\rho\rangle-\langle\gamma_m\Pi_A^+\phi,\tilde
P\rho\rangle\} dy \,.
\end{eqnarray*}

As we must replace $\gamma_m$ by $-\tilde\gamma_m$ in passing to
the dual structures, we have \begin{eqnarray*}
0&=&\beta_2(\phi,\rho,D,\BB)-\beta_2(\rho,\phi,\tilde
D,\tilde\BB)\\
&=&\textstyle\mathcal{C}(\phi,\rho,D,\BB)-\mathcal{C}(\rho,\phi,\tilde D,\tilde\BB) -\textstyle\int_M\{\langle D\phi,\rho\rangle-\langle\phi,\tilde D\rho\rangle\}dx \\
&&+\textstyle\int_{\partial M}\{-\langle\gamma_m\Pi_A^+P\phi,\rho\rangle-\langle\gamma_m\Pi_A^+\phi,\tilde P\rho\rangle -\langle\phi,\tilde\gamma_m\Pi_{A^{\#}}^+\tilde P\rho\rangle\\
&&\qquad-\langle
P\phi,\tilde\gamma_m\Pi_{A^{\#}}^+\rho\rangle\}dy\,.
\end{eqnarray*}
We now use the Greens formula given in Lemma \ref{lem-1.1} (3) to
see that $$
\mathcal{C}(\phi,\rho,D,\BB)=\mathcal{C}(\rho,\phi,\tilde
D,\tilde\BB)\,. $$ Thus we can assume that the integral
expressions for $\mathcal{C}$ are symmetric in $\phi$ and $\rho$.

If $\BB\phi=0$, then $\mathcal{C}(\phi,\rho,D,\BB)=0$ by Lemma
\ref{lem-2.1} (3). Similarly $\mathcal{C}(\phi,\rho,D,\BB)=0$ if
$\tilde\BB\rho=0$. Thus after eliminating divergence terms, the
integral formula for $\mathcal{C}$ is bilinearly expressible  in
terms of tangential operators applied to $$
\{\Pi_A^+\phi,\Pi_A^+P\phi\}\qquad\text{and}
\qquad\{\Pi_{A^{\#}}^+\rho,\Pi_{A^{\#}}^+\tilde P\rho\}\,.
$$

Since the boundary integrals defining $\mathcal{C}$ have total
weight $1$, terms which are bilinear in $\Pi_A^+P\phi$ and
$\Pi_{A^{\#}}^+\tilde P\rho$ do not appear. By Lemma \ref{lem-1.1}
(4), $$
\gamma_m\Pi_A^+=(\id-\tilde\Pi_{A^{\#}}^+)\gamma_m\quad\text{so}\quad
\textstyle\int_{\partial
M}\langle\gamma_m\Pi_A^+\Phi,\Pi_{A^{\#}}^+\tilde\Phi\rangle=0
$$
for any $\Phi$, $\tilde\Phi$. This shows that terms which are bilinear in $\Pi_A^+P\phi$ and $\Pi_{A^{\#}}^+\rho$ or in $\Pi_A^+\phi$ and $\Pi_{A^{\#}}^+\tilde P\rho$ do not involve $\gamma_m$. Taking into account the symmetry of $\mathcal{C}$, we see that these terms would have the form $$ \textstyle\int_{\partial M} b_0(\langle\Pi_A^+P\phi,\Pi_{A^{\#}}^+\rho\rangle+\langle\Pi_A^+\phi,\Pi_{A^{\#}}^+\tilde P\rho\rangle)dy\,. $$ Lemma \ref{lem-2.1} (5) now shows $b_0=0$. Consequently \begin{eqnarray*} &&\mathcal{C}(\phi,\rho,D,\BB)= \textstyle\int_{\partial M}\langle\mathcal{T}\Pi_A^+\phi,\Pi_{A^{\#}}^+\rho\rangle dy\quad\text{where}\\ &&\mathcal{T}=b_1A+b_2\gamma_mA+b_3A\gamma_m+b_4\gamma_mA\gamma_m\\
&&\hphantom{\mathcal{T}}+c_3L_{aa}\id+b_5\psi_P+b_6\gamma_m\psi_P
+b_7\psi_P\gamma_m+b_8\gamma_m\psi_P\gamma_m\\
&&\hphantom{\mathcal{T}}+b_9\psi_A+b_{10}\gamma_m\psi_A+b_{11}\widetilde{\psi_{A^{\#}}}\gamma_m
+b_{12}\widetilde{\psi_{A^{\#}}}\,.
\end{eqnarray*}
It is worth while making a few remarks about what invariants do
not appear. Modulo terms in $L_{aa}$, we can replace
$\gamma_m\psi_A\gamma_m$ by $\widetilde{\psi_{A^{\#}}}$ and
$\psi_A$ by $-\widetilde{\psi_{A^{\#}}}\gamma_m$. By  Lemma
\ref{lem-2.1} (5), $\gamma_mL_{aa}$ can not appear. Furthermore,
the invariants $\gamma_a\psi_P\gamma_a$ and
$\gamma_a\gamma_m\psi_P\gamma_a$ would violate Remark
\ref{rmk-2.5}.

Replacing $P$ by $\tilde P$ replaces $\gamma_m$ by
$-\tilde\gamma_m$, $A$ by $A^{\#}$, $\psi_P$ by $\tilde\psi_P$,
and $\psi_A$ by $\psi_{A^\#}$. Thus the symmetry of Lemma 2.1 (2)
yields $b_1=b_4$, $b_6=-b_7$, and $b_9=b_{12}$. Lemma 2.1 (5)
implies $b_2=b_3=b_5=b_8=b_{10}=b_{11}=0$. Setting  $b_1=b_4=c_2$,
$b_6=-b_7=c_4$, and $b_9=b_{12}=c_5$ then yields the formula of
the Lemma; we use Remark \ref{rmk-2.5} to see the coefficients
$c_i$ are universal. \end{proof}

\medbreak We complete the proof of Theorem \ref{thm-1.2} (3) by
showing: \begin{lemma}\label{lem-3.4} \begin{enumerate} \item
$c_2=\frac12$, $c_4=-\frac12$, $c_5=-\frac12$. \item
$c_3=\frac12$. \end{enumerate} \end{lemma}

\begin{proof} Again, we take the flat metric in Example \ref{exm-2.3}.
The flat connection is compatible with the Clifford module structure. It is not, however,
the only possible compatible connection. Let $\varrho_a$ be auxiliary real constants.
We define a compatible connection by setting $\omega_a:=\varrho_a\id$. As $\gamma_a$ and
$\gamma_0$ anti-commute with $\gamma_m$,
$$\begin{array}{lll} \psi_P=\delta_1\gamma_m\gamma_0-\varrho_a\gamma_a&\text{so}&
\gamma_m\psi_P-\psi_P\gamma_m=-2\delta_1\gamma_0-2\gamma_m\gamma_a\varrho_a,\\
\psi_A=\delta_2\gamma_0+\varepsilon\gamma_m\gamma_a\varrho_a&\text{so}&
\psi_A+\gamma_m\psi_A\gamma_m=2\delta_2\gamma_0+2\varepsilon\gamma_m\gamma_a\varrho_a\,.
\vphantom{\vrule height 11pt}
\end{array}$$
We take $\rho=0$ near $r=1$ so only the component where $r=0$ is
relevant in integrating over $\partial M$. On this component,
$\partial_r$ is the inward geodesic normal and we set
$\varepsilon=1$. By Lemma \ref{lem-3.2}, \begin{eqnarray*}
&&\beta_2(\phi,\rho,D,\BB)=-\textstyle\int_M\langle
D\phi,\rho\rangle
    +\textstyle\int_{\partial M}\{-\langle\gamma_m\Xi_+\gamma_m(\partial_r+\delta_1\gamma_0)\phi,\rho\rangle\\
&+&\langle\gamma_m\Xi_+\phi,\tilde\gamma_m(\partial_r+\delta_1\tilde\gamma_0)\rho\rangle
    +\langle((2c_2+2c_5)\delta_2\gamma_0\\
&+&(2c_5-2c_4)\gamma_m\gamma_a\varrho_a
-2c_4\delta_1\gamma_0)\Xi_+\phi,\tilde\Xi_+\rho\rangle\}dy\\
&=&-\textstyle\int_M\langle D\phi,\rho\rangle
+\textstyle\int_{\partial
M}\{\langle\Xi_-(\partial_r+\delta_1\gamma_0)\phi,\rho\rangle
   -\langle\Xi_+\phi,\partial_r\rho\rangle\\
&+&\langle[(-1-2c_4)\delta_1\gamma_0+(2c_2+2c_5)\delta_2\gamma_0
+(2c_5-2c_4)\gamma_m\gamma_a\varrho_a]\Xi_+\phi,\rho\rangle\}dy\,.
\end{eqnarray*}
On the other hand, since $P_0^2=-(\partial_r^2-\delta_1^2)\id$, the connection defined by $D_0$
is the trivial connection. Thus by Theorem \ref{thm-2.2}, \begin{eqnarray*}
&&(2\pi)^{m-1}\beta_2(\phi,\rho,D_0,\BB_0)\\
&=&-\textstyle\int_M\langle D\phi,\rho\rangle dx
+\textstyle\int_{\partial
M}\{\langle\Xi_-(\partial_r+\delta_1\gamma_0)\phi,\rho\rangle
-\langle\Xi_+\phi,\partial_r\rho\rangle\}dy\,.
\end{eqnarray*}
By Lemma \ref{lem-2.4}, $\beta_2(\phi,\rho,D,\BB)=
(2\pi)^{m-1}\beta_2(\phi,\rho,D_0,\BB_0)$. We may now complete the
proof of Assertion (1) by deriving the relations
$$2c_2+2c_5=0,\quad 2c_5-2c_4=0,\quad\text{and}\quad 2c_4=-1\,.$$

To study the coefficient of $L_{aa}$, we let $f$ be arbitrary in Example \ref{exm-2.3} but set $\delta_1=0$.
We have $\widetilde{A^{\#}}=A$; thus $\tilde\Pi_{A^{\#}}^+=\Pi_A^+$. Let $\phi=\phi(r)$ and $\rho=\rho(r)$.
We suppose $\rho$ vanishes identically near $r=1$ and suppress $\varepsilon$.
We may then apply Lemma \ref{lem-2.4} and Theorem \ref{thm-2.2} with $S=0$ to compute:
\begin{eqnarray*} &&\beta_2(\phi,g^{-1}\rho,D,\BB)=(2\pi)^{m-1}\beta_2(\phi,\rho,D_0,\BB_0)\\
&=&\textstyle\int_{M}-\langle D_0\phi,\rho\rangle
drd\theta+\textstyle\int_{\partial M}\{
\langle\Xi_-\partial_r\phi,\rho\rangle-\langle\Xi_+\phi,\partial_r\rho\rangle\}d\theta\,.
\end{eqnarray*}
We have $\Pi_A^+=\Xi_+$ and $\Pi_{A^{\#}}^+=\tilde\Xi_+$. Since
$\tilde Pg^{-1}\rho=-g^{-1}\tilde\gamma_m\partial_r\rho$ and
$dy=gd\theta$, \begin{eqnarray*} &&\textstyle\int_{\partial M}
\{\langle\Xi_-\partial_r\phi,\rho\rangle-\langle\Xi_+
\phi,\partial_r\rho\rangle\}d\theta\\
&=&\textstyle \int_{\partial
M}\{-\langle\gamma_m\Pi_A^+P\phi,g^{-1}\rho\rangle-\langle\gamma_m\Pi_A^+\phi,\tilde
Pg^{-1}\rho\rangle\}dy\,.
\end{eqnarray*}
Consequently we have
\begin{equation}\label{eqn-3.b}
0=\{c_2(A+A^{\#})+c_3L_{aa}+c_4(\gamma_m\psi_P-\psi_P\gamma_m)
+c_5(\psi_A+\widetilde{\psi_{A_{\#}}})\}\Pi_A^+\phi\,.
\end{equation}
We must define a compatible connection. Let
$\omega_m=0$ and $\omega_a=\textstyle\frac12\partial_rf\cdot\gamma_m\gamma_a$
define a connection $\nabla$ on $V$. We have \begin{eqnarray*} &&\Gamma_{mab}=\Gamma_{amb}
=-\Gamma_{abm}=\delta_{ab}e^{2f}\partial_rf,\\
&&\Gamma_{ma}{}^b=\Gamma_{am}{}^b=\delta_{ab}\partial_rf,\quad\text{and}\quad
\Gamma_{ab}{}^m=-\delta_{ab}e^{2f}\partial_rf\,.
\end{eqnarray*}
We have
$\gamma_{j;i}=\partial_i^x\gamma_j-\Gamma_{ij}{}^k\gamma_k+[\omega_i,\gamma_j]$.
We show $\nabla\gamma=0$ by checking:
\begin{eqnarray*}
&&\gamma_{m;m}=0,\\ &&\gamma_{a;m}=\partial_rf\cdot\gamma_a-\Gamma_{ma}{}^b\gamma_b=0,\\
&&\gamma_{m;a}=-\Gamma_{am}{}^b\gamma_b+[\omega_a,\gamma_m]=-\partial_rf\cdot\gamma_a
+\ffrac12\partial_rf[\gamma_m\gamma_a,\gamma_m]=0,\\
&&\gamma_{a;b}=-\Gamma_{ba}{}^m\gamma_m+[\omega_b,\gamma_a]
=\partial_rf\cdot
e^{2f}\gamma_m\delta_{ab}+\ffrac12\partial_rf[\gamma_m\gamma_b,\gamma_a]=0\,.
\end{eqnarray*}
Consequently
\begin{eqnarray*} &&\psi_P=-\gamma^a\omega_a=-\ffrac12\partial_rf\gamma^a\gamma_m\gamma_a=
-\ffrac12(m-1)\partial_rf\gamma_m,\\
&&\psi_A=
\gamma_m\gamma^a\omega_a+\delta_2\gamma_0=-\ffrac12(m-1)\partial_rf+\delta_2\gamma_0
\,.\end{eqnarray*} Thus $\gamma_m\psi_P-\psi_P\gamma_m=0$ and
$\psi_A+\widetilde{\psi_{A^{\#}}}=L_{aa}\id+2\delta_2\gamma_0$.
Furthermore as $\phi$ and $\rho$ are independent of $\theta$,
$A+\widetilde{A^{\#}}=2\delta_2\gamma_0$. Thus Equation
(\ref{eqn-3.b}) yields $c_3+c_5=0$ so, by Assertion (1),
$c_3=-c_5=\frac12$.  This completes the proof of Lemma
\ref{lem-3.4} and thereby the proof of Theorem \ref{thm-1.2}.
\end{proof}

\section*{Acknowledgments}
Research of P. Gilkey partially supported by the Max Planck
Institute for the Mathematical Sciences (Leipzig). Research of K.
Kirsten partially supported by the Max Planck Institute for the
Mathematical Sciences (Germany) and the Baylor University Summer
Sabbatical Program.  Research of JH. Park supported by
R04-2000-00002 from the Korea Science and Engineering Foundation.

\end{document}